\documentclass[twocolumn, superscriptaddress, nofootinbib, showpacs,
amsmath, amssymb]{revtex4}

\begin{document}

\title{ Quantum-gravitational corrections to the hydrogen atom and harmonic
oscillator }

\author{ Michael~Maziashvili}
\email{mishamazia@hotmail.com} \affiliation{ Andronikashvili
Institute of Physics, 6 Tamarashvili St., Tbilisi 0177, Georgia
}\affiliation{ Faculty of Physics and Mathematics, Chavchavadze
State University, 32 Chavchavadze Ave., Tbilisi 0179, Georgia }
\author{ Zurab~Silagadze}
\email{Z.K.Silagadze@inp.nsk.su}\affiliation{ Budker Institute of
Nuclear Physics and Novosibirsk State University, Novosibirsk 630
090, Russia }

\begin{abstract}

It is shown that the rate of corrections to the hydrogen atom and
harmonic oscillator due to profound quantum-gravitational effect
of space-time dimension running/reduction coincides well with
those obtained by means of the minimum-length deformed quantum
mechanics. The rate of corrections are pretty much the same within
the accuracy by which we can judge the quantum-gravitational
corrections at all. Such a convergence of results makes the
concept of space-time dimension running more appreciable. As a
remarkable distinction, the energy shift due to dimension
reduction has the opposite sign as compared with the correction
obtained by means of the minimum-length modified quantum
mechanics. Thereby, the sign of total quantum-gravitational
correction remains obscure.

\end{abstract}






\maketitle

\section{\textsf{ Introduction  }}

Along the development of various scenarios for quantum gravity
(QG), the study of QG phenomenology becomes important. One of the
first intensive streams for such phenomenological study was
triggered by treating the general relativity as an effective field
theory \cite{Donoghue}. This effective field theory approach, that
is, to consider general relativity as an effective low energy
approximation to some as yet unknown fundamental theory of quantum
gravity, offers a way to get round the familiar renormalization
difficulties of general relativity in the low energy regime
\cite{Donoghue}. Using this approach one can make reliable
predictions about the radiative QG corrections to various physical
quantities in the low energy limit ($E \ll E_P$) \cite{Donoghue,
Burgess}.

Another inspiration for a systematic study of QG phenomenology was
the minimum-length modified quantum mechanics \cite{KMM} that
stems from the generalized uncertainty relations. Generalized
uncertainty relations naturally arise in string theory
\cite{String} as well as in some heuristic QG considerations
\cite{heuristic}.

A new door for a versatile study of the QG phenomenology is open
by a profound QG effect of space-time dimension running/reduction
discovered recently in two different approaches to quantum gravity
\cite{AJLLR}. In this paper we study the QG corrections to the
hydrogen atom and harmonic oscillator due to dimension running and
compare the results with those obtained by means of the
minimum-length modified quantum mechanics. Throughout the paper we
assume the system of units $c = 1$. The paper is organized as
follows. First, in sections II and III we consider two distinct
qualitative approaches for estimating of QG energy shift to the
hydrogen atom that gives good understanding of the order of
magnitude for this correction. Then, in section IV we review the
results obtained for hydrogen atom in the framework of
minimum-length modified quantum mechanics. In this approach
different results were reported for the hydrogen atom that
reflects the technical difficulty of using the KMM quantum
mechanics \cite{KMM} in practice. So the further detailed elucidating
study of these results is in order. Next, in section V we provide a
simple consideration of the effect of QG dimension
running/reduction without resorting to any particular model of QG,
but rather on the basis of a finite space-time resolution that is
implied by all GQ scenarios. Throughout this consideration a
simple analytic expression of running dimension emerges. Using
this expression of running dimension we estimate the corrections
to the hydrogen atom and also consider how the dimension running
affects the relativistic and QED radiative corrections. From the
very outset let us notice that comparing all approaches considered
throughout this paper, an unique result emerges for the rate of QG
energy shift to the hydrogen atom, but the signs of corrections
that come in the framework of minimum-length modified quantum
mechanics and due to dimension running respectively, are opposite.
Further, in section VI we consider in a similar manner the
corrections to the harmonic oscillator due to minimum-length
modified quantum mechanics and the dimension running. Then follows
the concluding remarks.

\section{\textsf{ Semi-classical treatment }}

The hydrogen like atom has the well known non-relativistic
spectrum

\[ E_n = - {E_{R} \over n^2} ~,~~~~~~\mbox{where}~~~~~~E_{R} =
{Z\,^2e^4m \over 2\hbar^2}~,\]
with the orbital radii \[ a_n = n^2 a~,~~~~~~\mbox{where}~~~~~~a =
{\hbar^2 \over Zme^2}~. \] Throughout the paper we will tacitly
assume that the condition of non-relativistic motion of the
electron $e^2Z \ll n$ is satisfied. As is well known \cite{sctreatment},
we can successfully estimate the ground state parameters of the hydrogen
atom simply by using the position-momentum uncertainty relation
\[\delta x\delta p \ge {\hbar \over 2 }~.\] Certainly, all that
should be expected in the framework of this semiclassical
treatment is an order-of-magnitude estimate.\footnote{ Let us
notice that by taking into account that the radial wave function
with the principal quantum number $n$ has $(n-1)$ nodes and
therefore the electron in this state can be considered as
localized in the spatial region $\delta x\sim r/n$, that is, $p
\sim (\hbar n)/r$ we will get exact result for hydrogen spectrum
\cite{Kuo}.} Namely, the minimum of hydrogen energy
\begin{equation}\label{hydrogenen} E = {p^2 \over 2m} -
{Ze^2 \over r} ~, \end{equation} where on the grounds of
uncertainty relation $p$ is understood as $p = \hbar/r$, occurs at
$r = a$ that results in the ground state energy $E=-E_R$.
Following this line of discussion let us see how the ground state
parameters will change for QG modified uncertainty relation
\cite{KMM}
\[ \delta x \ge {\hbar \over 2\delta p} + {(\delta x_{min})^2
\over 2 \hbar }\,\delta p ~,\] which results from the commutation
relation
\[ [ \hat{x}, \hat{p} ] = i\hbar (1 + \beta \hat{p}^2) ~,
~~~~~~\mbox{where}~~~~~~ \delta x_{min} =
\hbar \sqrt{\beta} ~.\] Solving for the momentum uncertainty in
terms of the distance uncertainty, the equation \[\delta x =
{\hbar \over 2\delta p} + {(\delta x_{min})^2 \over 2 \hbar
}\,\delta p ~,
\] gives \[\delta p \,= \, \hbar \,\, {\delta x - \sqrt{(\delta x)^2 -
(\delta x_{min})^2} \over (\delta x_{min})^2} \,\approx \, {\hbar
\over 2\delta x} \, + \, {\hbar \,(\delta x_{min})^2 \over 8
\,(\delta x)^3}~.\] Now minimizing the hydrogen energy
(\ref{hydrogenen}) with this expression of momentum, $p=2\delta
p$, where $\delta x$ is replaced by $r$,  from ${dE / dr} = 0$ one
finds
\[ {mZe^2 \over r^2} - \left({\hbar \over r} + {\hbar \,(\delta x_{min})^2
\over 4\, r^3} \right)\left({\hbar \over r^2} + {3\hbar
\,(\delta x_{min})^2 \over 4\, r^4} \right) =0 ~.\] To the lowest
order in $\delta x_{min}$ this equation reduces to \[ mZe^2r^3 -
\hbar^2  r^2  -  \hbar^2 (\delta x_{min})^2  =
0~.\] The solution of this equation to the lowest order in $\delta
x_{min}$ looks like \[ r = {\hbar^2 \over mZe^2} \, +\,
{mZe^2(\delta x_{min})^2 \over \hbar^2} = a  \,+\,
{(\delta x_{min})^2 \over a}~,\] that leads to the ground state
energy
\[ E = - E_R \left[ 1 - \frac{1}{2} \left({\delta x_{min} \over a }\right)^2
\right]~.\] So from minimum-length modified quantum mechanics we
get $\delta E \sim E_R\left(\delta x_{min}/ a\right)^2$.

\section{\textsf{ One more qualitative treatment }}

{\bf \textsf{ Instructive example from QED. --}} In early days of
the systematic treatment of QED infinities, Welton gave a simple
qualitative description of the Lamb-Retherford effect \cite{LR} by
considering the interaction of electron with the vacuum
fluctuations of the electromagnetic field \cite{QED, Khriplovich}.
Upon treating the electron classically and non-relativistically,
the equation governing the fluctuations of the electron takes the
form
\begin{equation}
\label{flucteq}
m\frac{d^2(\delta \mathbf{r})}{dt^2} = e\,  \boldsymbol{\mathcal{E}}\,,
\end{equation}
where $\mathcal{E}$ denotes the fluctuating electric field. In its
vacuum state, the electromagnetic field is characterized by the
mean-square fluctuations attributed to the zero-point
oscillations,
\[ \frac{\left\langle \mathcal{E_\omega}^2 \right\rangle }{4\pi}=
\frac{\hbar\,\omega}{2}~. \]
Under the influence of a given Fourier component of the electric field
$\boldsymbol{\mathcal{E}}$ with the frequency  $\omega$, electron is subject
to the oscillations which have the amplitude
\[\delta r_{\omega} = -\frac{e\mathcal{E}_{\omega}}{m\omega^2}~.\]
For the mean-square fluctuations one finds $\left\langle (\delta
r_{\omega})^2 \right\rangle = e^2 \langle \mathcal{E}_{\omega}^2
\rangle / m^2 \omega^4 $ and, respectively,
\begin{equation}
\label{elposfluct} \left\langle (\delta r)^2\right\rangle =
2\frac{e^2}{m^2}\int \frac{\left\langle \mathcal{E}_{\omega}^2
\right\rangle} {\omega^4}\,\frac{d^3 k}{(2\pi)^3} =
\frac{2e^2\hbar}{\pi m^2} \int \frac{d\omega}{\omega}~,
\end{equation}
where we have summed over polarizations as well. To give a
concrete meaning to this formal expression, we need to specify the
integration limits for the frequency integral (\ref{elposfluct}).
As the system under consideration has the binding energy $\sim
(Z^2 e^4 m)/\hbar^2$, the natural lower limit on the fluctuation
frequencies is set up by this binding energy, because for lower
frequencies the electron can not be considered as a free and the
equation (\ref{flucteq}) becomes invalid. On the other hand, as
the electron can not be probed beneath its Compton wavelength,
there is a natural upper limit on the integration set by the
electron mass. Hence we get
\[ \label{elfluct}\left\langle (\delta
r)^2\right\rangle \simeq  \frac{2e^2\hbar}{\pi m^2}
\int\limits_{(e^4 Z^2 m)/\hbar^3} \limits^{m/\hbar}
\frac{d\omega}{\omega} = \frac{2e^2\hbar}{\pi m^2}
\ln{\frac{\hbar^2}{e^4Z^2}}~.\] As the electron is forced to
fluctuate around the equilibrium position, it sees the Coulomb
potential to be somewhat smeared out. The second term of the
Taylor expansion
\[V(\mathbf{r}+\delta\mathbf{r}) =
V(\mathbf{r}) + \mathbf{\nabla} V\cdot \delta\mathbf{r} + \frac{1}{2}
\sum\limits_{i,\,j} \delta r^i\delta r^j \frac{\partial^2V}
{\partial r^i\partial r^j}+ \cdots~,\]
gives the zero average effect. Therefore, the first non-trivial correction
has the form
\[\delta V = \frac{1}{6} \left\langle (\delta r)^2\right\rangle \triangle
\frac{-e^2Z}{r} = \frac{2\pi Ze^2}{3}\,\left\langle (\delta
r)^2\right\rangle\,\delta(\mathbf{r}) ~.\] The average effect of
this smearing of the potential in a given eigenstate of the atom
will result in the energy shift
\[\delta E_n = \int \delta V \left|\psi_n(\mathbf{r})\right|^2d^3r
= \frac{2\pi Ze^2}{3}\,\left\langle (\delta r)^2\right\rangle
\left|\psi_n(0)\right|^2~.\] The wave function at the origin
$\left|\psi_n(0)\right|$ vanishes for all states with non-zero
angular momentum. For $l = 0$, after inserting in this equation
the values of $\left\langle (\delta r)^2\right\rangle$ and
$\left|\psi_n(0)\right|^2 = Z^3m^3e^6/\pi n^3\hbar^6$ we get
\[
\delta E_n \simeq \frac{8mZ^4 e^{10}}{3\pi
n^3\hbar^5}\ln{\frac{\hbar} {e^2Z}}~.\] This result accounts well
for the experimental observations. In fact, for $l \neq 0$ states
the experimentally measured energy shifts are not precisely zero,
but they are much smaller than the energy shift of the $2s_{1/2}$
level.

{\bf \textsf {Applying QG setup. --}} In the case of QG treatment,
the situation is somewhat simplified as we know $\delta r$ from
the very outset. Namely, in QG the position of electron can not be
specified better than $\delta x_{min}$, thus for position
fluctuation of the electron one simply finds $\delta r = \delta
x_{min}$. For $l = 0$, the QG induced energy shift takes the form
\[\delta E_n \simeq \frac{2}{3}Ze^2(\delta x_{min})^2
\left( \frac{Zme^2}{n\hbar^2} \right)^3 =\frac{4}{3}
\frac{E_R}{n^3} \left( \frac{\delta x_{min}}{a} \right)^2~.\] We
see the correction is of the same order as that one considered in
previous section.

\section{\textsf{ Minimum-length modified quantum mechanical treatment }}

When the system has several degrees of freedom the minimum-length
modified commutation relation considered in section II generalizes
to \cite{KMM}
\begin{equation}\label{minlengthqm} [\hat{x}_i,\hat{p}_j]=i\hbar
\left(\delta_{ij}+\beta\hat{p}^2\delta_{ij} +
\beta^\prime\hat{p}_i\hat{p}_j \right)~.\end{equation} The minimum
length which follows from these commutation relations is (for more
details see \cite{KMM,Kempf})
\[\delta x_{min}=\hbar\sqrt{3\beta+\beta^\prime}~.\] In a particular case
$\beta^\prime=2\beta$, i.e.,
\begin{equation}\label{partcase}
[ \hat{x}_i, \hat{p}_j ] = i\hbar( \delta_{ij}
          + \beta \hat{p}^2 \delta_{ij}
          + 2\beta \hat{p}_i \hat{p}_j
        )~,
\end{equation} the realization of this algebra to the linear order in
$\beta$ can be done in a simple way in terms of the standard
position and momentum operators $ \left[ \hat{x}^0_i, \hat{p}^0_j
\right] = i\hbar \delta_{ij}$ \cite{Brau}. Indeed, defining
\begin{equation}\label{brauapproach} \hat{x}_i = \hat{x}^0_i~,~~
~~ \hat{ p}_i = \hat{ p}^0_i \left[ 1 + {\beta} \left(\hat{\mathbf
p}^0\right)^2 \right] ~,\end{equation} one easily finds that to
the linear order in $\beta$ these operators satisfy the algebra
(\ref{partcase}) \cite{Brau}. Working to this accuracy one gets
the following universal QG correction to the Hamiltonian
\cite{Brau}
\begin{equation} \hat{\mathcal{H}}  =  {\hat{\mathbf p}^2 \over 2m} +
V(\hat{\mathbf r}) = {\left(\hat{\mathbf p}^0\right)^2 \over 2m} +
V(\hat{\mathbf r}^0) + {\beta \over m}\left(\hat{\mathbf
p}^0\right)^4 + O(\beta^2)~.
\end{equation}
Written in this way, the only appeal to the
dynamical system under consideration is to supply us with the
Hamiltonian $\hat{\mathcal{H}}^0$. Denoting by $\psi^0_n,\, E^0_n$
the eigenfunctions and eigenvalues of the unperturbed operator
$\hat{\mathcal{H}}^0$
\begin{equation}\label{eigeneq}  \hat{\mathcal{H}}^0 \psi^0_n \equiv
\left[ {\left(\hat{\mathbf p}^0\right)^2 \over 2m} +
V(\hat{\mathbf r}^0) \right ]\psi^0_n = E^0_n\psi^0_n ~,
\end{equation} for the first order correction to the eigenvalue
$E^0_n$ one finds \cite{LL}
\[ \delta E_n =  {\beta \over m} \left \langle \psi^0_n \right|
\left(\hat{\mathbf p}^0\right)^4\left| \psi^0_n \right\rangle ~.\]
From now on we will work in the coordinate representation
$\hat{\mathbf r}^0 =  {\mathbf r},\, \hat{\mathbf p}^0 = -i\hbar
\partial_{{\mathbf r}}$. Using the Eq.(\ref{eigeneq}) and the fact that
$\hat{\mathbf p}^0$ is a Hermitian operator, one finds
\begin{widetext}
\begin{eqnarray}\label{encorrec}
{\beta \over m} \left\langle \psi^0_n \right|\left(\hat{\mathbf
p}^0\right)^4\left| \psi^0_n \right\rangle & = & 2\beta \left
\langle \psi^0_n \right|\left(\hat{\mathbf p}^0\right)^2 \left[
E^0_n - V({\mathbf r}) \right ]\left|\psi^0_n\right \rangle
 \nonumber\\ & = & 2\beta
\left \langle \left(\hat{\mathbf p}^0\right)^2 \psi^0_n \right|
\left[ E^0_n - V({\mathbf r}) \right ]\left|\psi^0_n \right
\rangle   = 4\beta m \left \langle  \psi^0_n \right| \left[ E^0_n
- V({\mathbf r}) \right ]^2\left|\psi^0_n \right \rangle~.
\end{eqnarray}
\end{widetext} From Eq.(\ref{encorrec}), for the Hydrogen atom we
get \cite{Brau}
\begin{eqnarray}\label{hydrogen} \delta E_{n\ell} & = &  {\beta \over m}
\left \langle \psi^0_{n\ell} \right|\left(\hat{\mathbf p}^0\right)^4\left|
\psi^0_{n\ell} \right\rangle
\nonumber \\ & = & 4\beta m \int\limits_0\limits^{\infty}dr\,
r^2R_{n\ell}^2 \left[ {mZ^2e^4 \over 2\hbar^2 n^2} - {Ze^2 \over
r} \right ]^2~,\end{eqnarray} where $\psi^0_{n\ell} =
R_{n\ell}(r)Y_{\ell m}(\vartheta,\,\varphi)$ are eigenfunctions
for the Coulomb problem \cite{LL}.

Using the Bohr radius $a=\hbar^2/mZe^2$ as a length unit, one can
write the Eq.(\ref{hydrogen}) in the form
\begin{equation}\label{gravencorrhyd}\delta E_{n\ell} = 4\beta m
\left({mZ^2e^4 \over \hbar^2}\right)^2
\int\limits_0\limits^{\infty}d\xi\, \xi^2 \tilde{R}_{n\ell}^2 \left[ {1
\over 2 n^2} - {1 \over \xi} \right ]^2 ~,\end{equation} where
$$\xi=\frac{r}{a},\;\;\;\tilde{R}_{n\ell}=a^{3/2}R_{n\ell}.$$
Now the integrand is written in dimensionless quantities and
$\tilde{R}_{n\ell}$ function has an explicit form \cite{LL}
\[\tilde{R}_{n\ell} = -{2 \over n^2} \sqrt{{(n-\ell-1)! \over
\left[(n+\ell)! \right]^3}}\,e^{-\xi/n}\left( {2\xi \over n}
\right)^{\ell}L^{2\ell +1}_{n+\ell}\left( {2\xi \over n}
\right)~.\]
Using the normalization condition for $\tilde{R}_{n\ell}$
$$\int\limits_0^\infty \tilde{R}^2_{n\ell}(\xi)\,\xi^2\,d\xi=1,$$
and the well known mean values \cite{LL,Qiang}
$$\overline{\xi^{-1}}=\frac{1}{n^2},\;\;\;\overline{\xi^{-2}}=\frac{1}
{n^3\left (\ell+\frac{1}{2}\right )},$$
where
$$\overline{\xi^k}=\int\limits_0^\infty \tilde{R}^2_{n\ell}(\xi)\,
\xi^{2+k}\,d\xi,$$
we reproduce the result of \cite{Brau}:
\begin{equation}\label{Brau} E_{n\ell}=-E_R\left [
\frac{1}{n^2}-2\,\frac{4n-3(\ell + 1/2)}{5(\ell +
1/2)}\left(\frac{\delta x_{min}} {a_n}\right )^2
\right]~.\end{equation}
Therefore the QG correction works with a positive sign.

The use of KMM quantum mechanics \cite{KMM} that exactly satisfies
the modified commutation relations (\ref{minlengthqm}) is
technically more complicated. Using KMM quantum mechanics, the
posterior study of the hydrogen atom led authors of the paper
\cite{AkhouryYao} to the following result for $\ell = 0$ energy
levels
\begin{equation}\label{AkhouryYao} E_n=-\frac{E_{R}}{n^2}\left[1 +
2(\beta+\beta')\frac{m^2Z\,^2
e^4}{\hbar^2n^2}\right]~,\end{equation}
implying in the case $\beta^\prime=2\beta$
\[\delta E_n \simeq -\frac{6}{5}E_{R}\left(\frac{
\delta x_{min}}{a_n } \right)^2 ~.\] Albeit the rate of the
correction is about the same the sign is opposite as compared with the
Eq.(\ref{Brau}). The subsequent consideration of the hydrogen atom
in the KMM formalism \cite{BChMT}, exhibits positive QG correction
in agreement with the Eq.(\ref{Brau}) but unfortunately this paper
could neither account for the discrepancy nor reproduce the
results of Eqs.(\ref{Brau},\,\ref{AkhouryYao}). Further study for
a final clarification of the sign of QG correction in the KMM
approach is desired.

It is noteworthy that in the KMM construction the inverse distance
operator $\hat r^{-1}$ is non-local \cite{KMM} and, therefore, in
this approach the Coulomb potential smearing considered in section
III is automatically taken into account. While the above described
scheme for an approximate realization of the minimum-length
modified commutation relations, Eq.(\ref{brauapproach}), misses
this effect.

\section{\textsf{ Hydrogen atom in view of the QG running/reduction of
space-time dimension }}

\subsection*{\textsf{ QG running/reduction of space-time dimension }}

Because of quantum gravity the dimension of space\,-\,time appears
to depend on the size of region, it is somewhat smaller than four
at small scales
and monotonically increases with increasing the size of the region
\cite{AJLLR}. We can account for this effect in a simple and
physically clear way that allows us to write simple analytic
expression for space\,-\,time dimension running. Let us consider a
subset $\mathcal{F}$ of four dimensional Euclidean space
$\mathbb{R}^4$, and let $l^4$ be a smallest box containing this
set, $\mathcal{F} \subseteq l^4$. For estimating the dimension of
$\mathcal{F}$ we have to cover it by $\epsilon^4$ cells and
counting the minimal number of such cells, $N(\epsilon)$, we
determine the dimension, $d \equiv \dim(\mathcal{F})$ as a limit
$d = d(\epsilon \rightarrow 0)$, where $n^{d(\epsilon)} = N $ and
$n = l/\epsilon$. For more details see \cite{Falconer}. This
definition is referred to as a box-counting dimension and can be written
in a more familiar form as
\[ d =\lim\limits_{\epsilon \rightarrow 0} {\ln N(\epsilon) \over
\ln {l \over \epsilon}}~.\]
Certainly, in the case when $\mathcal{F} = l^4$, by taking the
limit $d(\epsilon \rightarrow 0)$ we get the dimension to be $4$.
From the fact that we are talking about the dimension of a set
embedded into the four dimensional space, $\mathcal{F} \subset
\mathbb{R}^4$, it automatically follows that its dimension can not
be greater than $4$, $d \le 4$. We see that the volume of a
fractal $\mathcal{F}$ uniformly filling the box $l^4$ is reduced
\[ V(\mathcal{F}) = \lim\limits_{\epsilon \rightarrow 0} N(\epsilon)\epsilon^4
= \lim\limits_{\epsilon \rightarrow
0}n(\epsilon)^{d(\epsilon)}\epsilon^4 ~,\] in comparison with the
four dimensional value $l^4$. Introducing $\delta N =
n(\epsilon)^4 - N(\epsilon)$, the reduction of dimension
$\varepsilon = 4 - d$ can be written as
\begin{equation}\label{opdim} \varepsilon(\epsilon) \,=\, - {\ln
\left(1 - {\delta N(\epsilon) \over n(\epsilon)^4 } \right)\over
\ln n(\epsilon) } \, \approx \, {1 \over \ln n(\epsilon)}
\,{\delta N(\epsilon) \over n(\epsilon)^4}~. \end{equation} In
quantum gravity the space\,-\,time resolution is set by the
minimum length $\epsilon = \delta x_{min}$. The local fluctuations
$\sim \delta x_{min}$ add up over the length scale $l$ to $\delta
l = (\delta x_{min}l)^{1/2}$ \cite{mazia1}. Respectively, for the
region $l^4$ we have the deviation (fluctuation) of volume of the
order $\delta V = \delta l^4$, where $\delta l$ depends on the
scale $l$ as indicated above. Thus in quantum gravity we expect
the Poison fluctuation of volume $l^4$ of the order $ \delta V =
(l^2/\delta x_{min}^2)\, \delta x_{min}^4 $ \cite{Sorkin}. One
naturally finds that this fluctuation of volume has to account for
the reduction of dimension.\footnote{ This suggestion has been
made in \cite{mazia1}, though the rate of volume fluctuation was
overestimated in this paper, see \cite{mazia2}. Let us also notice
that the necessity of operational definition of dimension because
of quantum mechanical uncertainties (not quantum\,-\,gravitational
!) was first stressed in \cite{SZ}.} Respectively, $n = l/\delta
x_{min}$, $\delta N = l^2/\delta x_{min}^2$, and from
Eq.(\ref{opdim}) one gets,
\[ \varepsilon = {1 \over \ln {l\over \delta x_{min}}}\,
\left({\delta x_{min} \over l}\right)^2~.\]
This equation gives the running of dimension with
respect to the size of region $l$.

\subsection*{\textsf{  Hydrogen atom in light of the QG reduction of
dimension }}

One can analytically solve the Schr\"odinger equation for Coulomb
potential in arbitrary $D$ spatial dimensions, that gives \cite{BLM,He}

\[\widetilde{E}_n = - \frac{E_{R}}{\left(n+ \frac{D - 3}
{2}\right)^2}~,\] and the orbital radii take the form
\[ \widetilde{a}_n = \left(n+ \frac{D - 3 }{2}\right)^2 a~.\]
For $D = 3 - \varepsilon$, where  $\varepsilon \ll 1$ one finds
\[ \widetilde{E}_n = -E_{R}\left (\frac{1}{n^2} + \frac{\varepsilon}{n^3}
\right) ~,~~~~~~ \widetilde{a}_n = n^2a -  \varepsilon na ~.\]
Inserting the above found expression for QG dimension
running/reduction
\[ \varepsilon =  \frac{1}{\ln{ \frac{a_n}{\delta x_{min}}}}\,
\left(\frac{\delta x_{min}}{a_n}\right)^2~,\] we find
\[ \delta E_n \simeq -
\frac{E_R}{n^3\ln \frac{a_n}{\delta x_{min}}}\left(\frac{\delta
x_{min}}{a_n}\right)^2\,.\] Apart from the less important
numerical (logarithmic) factors, the rate of correction $\sim
E_R(\delta x_{min}/a)^2$ is in perfect agreement with the results
of previous sections. Indeed, the factor $\ln(l/\delta x_{min})$
is of less significance as for the length scale $l$ at which the
QG corrections become important the ratio $l/\delta x_{min}$ is
not very large. So, in view of the precision we can pretend to in
study of the QG corrections this factor is really less important.
We see that the correction due to QG dimension reduction works
with a negative sign, that is, it increases the binding energy.

The angular momentum eigenvalues also change because of dimension
reduction \cite{He}
\[L = \sqrt{\ell(\ell+ D - 2)}\,\hbar  \,\approx \sqrt{\ell(\ell+ 1)}\,\hbar  -
\frac{\varepsilon \hbar \,\ell}{2\sqrt{\ell(\ell+ 1)}} \,~.\]

It should be noticed that in this discussion we have not taken
into account the modification of Coulomb potential $\sim 1/r$ due
to dimension reduction $\sim 1/r^{D-2}$.

\subsection*{\textsf{ Shift of radiative and relativistic corrections due to
dimension reduction}}

The  dimensional regularization approach allows one to simply
estimate how the dimension reduction affects the well known QED
radiative corrections \cite{SZ,Nakamura, Svozil}. Namely, in
calculating QED radiative corrections to the photon propagator or
to the photon-electron vertex, we have due to integration in
momentum space the well known $\Gamma(\varepsilon/2)$ factor (see,
for example \cite{QFT}). Now, keeping simply the terms linear in
$\varepsilon$ in the radiative corrections, that come from the
decomposition of the $\Gamma(\varepsilon/2)$ function alluded to
above,
\[\Gamma\left(\frac{\varepsilon}{2}\right) = \frac{2}{\varepsilon} -
\gamma + \frac{\varepsilon}{4} \left(\frac{\pi^2}{6} + \gamma^2\right) +
O(\varepsilon^2)~,~~~~ \gamma\approx 0.5772,\] one easily infers
that due to dimension reduction we will have the shift of the
standard QED radiative corrections which are of the order
\[  \sim \,\varepsilon \times \mbox{QED radiative corrections}~.\]
In this way one finds that the corrections due to QG dimension
reduction to the electron anomalous magnetic moment and the Lamb
shift are of the order of $\sim\varepsilon e^3/m$ and $ \sim
\varepsilon Z\,^2e^6E_R/\hbar^3n^3$ respectively, see
\cite{Svozil}. For this energy shift we see that because of factor
$e^6/\hbar^3 \simeq (1/137)^3$ it is by about six orders of
magnitude smaller than the leading QG correction $\sim\varepsilon
E_R/n^3$ found above.

To see how the dimension reduction affects the relativistic
corrections to the hydrogen energy levels, one can use formally
the solution of the Dirac equation in $D+1$ space-time dimensions
for the Coulomb potential \cite{D-Dirac}
\begin{equation}\label{DDirac} \widetilde{E}_n = m \left(1 +
\frac{(Ze^2/\hbar)^2}{\left(\sqrt{K^2-(Ze^2/\hbar)^2} +n -\ell -1
\right)^2 } \right)^{-1/2} ~,\end{equation} where $K = (2\ell + D
- 1)/2$ and assume $D = 3 -\varepsilon$. Expanding
Eq.(\ref{DDirac}) in powers of $(Ze^2/\hbar)^2$, one finds

\begin{eqnarray}  \widetilde{E}_n &=&  m \,-\, \frac{m(Ze^2/\hbar)^2}
{2\left(n+\frac{D-3}{2}\right)^2} \nonumber \\ &-&
\frac{m(Ze^2/\hbar)^4}{2\left(n+ \frac{D-3}{2}\right)^4}\left(
\frac{2n+D-3}{2\ell + D-1} - \frac{3}{4}\right)  ~,\nonumber
\end{eqnarray} where the first term on the right-hand side represents the rest
energy $m$, the second term, which was discussed above, comes from
the Schr\"odinger equation, and the third one describes
relativistic corrections. Substituting $D = 3 - \varepsilon$,
where $\varepsilon$ is estimated at the atomic scale $l = a_n$,
one easily finds the shift of relativistic corrections due to
dimension reduction. This shift of relativistic correction due to
dimension reduction is suppressed in comparison with the leading
QG correction $\sim\varepsilon E_R/n^3$ by the factor
$(Ze^2/\hbar)^2$, that is by about four orders of magnitude
for hydrogen atom.

Again, the modification of the Coulomb potential because of
dimension reduction $1/r \rightarrow  1/r^{D-2}$ is ignored
throughout this discussion.

\section{\textsf{  Harmonic oscillator  }}

Assuming the modified commutation relations (\ref{minlengthqm}),
for $D$-dimensional harmonic oscillator
\[
\hat{\mathcal{H}} = {\hat{\mathbf p}^2 \over 2m}
        +  {m\,\omega^2
\hat{\mathbf r}^2 \over 2}~,
\] one finds \cite{KChMOT}
\begin{widetext}
\begin{eqnarray}
E_{n\ell} & = & \hbar\omega \left[
  \left( n + \frac{D}{2} \right)
        \sqrt{ 1 + \left\{ {\beta^2 L^2 \over \hbar^2 }
                         + \frac{ (D\beta + \beta')^2 }{ 4 }
                   \right\}m^2\hbar^2\omega^2
             }
\right. \cr && \qquad\left. + \left\{ (\beta + \beta')\left( n +
\frac{D}{2}\right)^2
        + (\beta - \beta')\left({L^2\over \hbar^2} + \frac{D^2}{4}\right)
        + \beta'\frac{D}{2}
  \right\}\frac{m\hbar\omega}{2}
\right]~, \nonumber
\end{eqnarray}
\end{widetext}
where $ L^2 = \ell (\ell + D -2 )\hbar^2$. With this expression
one easily considers the minimum-length modified quantum
mechanical correction to the harmonic oscillator against the
correction coming from QG reduction of dimension. For $D=3$, the
energy spectrum to the lowest order in $\beta,\,\beta'$ takes the
form
\begin{eqnarray}
E_{n\ell}  &\approx &\hbar\omega \left[ \left(n+\frac{3}{2}\right)
+ \frac{1}{2}
  \left\{ (k^2 + {k'}^2) \left(n+\frac{3}{2}\right)^2
       \right.\right. \nonumber \\&& \left.\left.  + (k^2 - {k'}^2) \left(\ell(\ell+1) +\frac{9}{4}\right)
        + {k'}^2\frac{3}{2}
  \right\}
\right] ~,
\end{eqnarray} where
\[
k^2 = \beta\hbar m\omega\,,\qquad {k'}^2 = \beta'\hbar m \omega ~.
\] To estimate the QG corrections due to dimension
running/reduction \[ E_{n}  =  \hbar\omega
  \left( n + \frac{D}{2} \right) = \hbar\omega
  \left( n + \frac{3}{2} \right) - \hbar\omega\frac{\varepsilon}{2} ~,\]
first we have to evaluate the size of a localization region for the system.
One can do this in a simple semiclassical way by equating
\[{m\,\omega^2 r^2 \over 2} = \hbar\omega
  \left( n + \frac{3}{2} \right)~~ \Rightarrow ~~r_n = \left[\frac{2\hbar}
{m\omega} \left( n + \frac{3}{2} \right)\right]^{1/2}~,\] For $\varepsilon$
one finds \[ ~~~~ \varepsilon = \frac{1}{\ln{ \frac{r_n}{\delta x_{min}}}}\,
\left(\frac{\delta x_{min}}{r_n}\right)^2~,\] where
\[ \left(\frac{\delta x_{min}}{r_n}\right)^2 = \frac{\hbar m\omega
(3\beta + \beta')}{2\left( n + \frac{3}{2} \right)}~.\] We see that
as in the case of hydrogen atom, the QG corrections to the
harmonic oscillator due to dimension reduction and minimum-length
modified quantum mechanics have the same rate up to less important
logarithmic factors but work with opposite signs. QG correction to the
harmonic oscillator due to dimension reduction works with negative sign as
in the case of hydrogen atom.

\section{\textsf{  Concluding remarks  }}

The qualitative treatments considered in sections II and III give
us a reliable idea about the rate of QG correction to the energy
spectrum of the hydrogen atom but probably tell little about the
proper sign of the correction. From the physical point of view,
one can identify three distinct sources of the QG corrections to
the hydrogen atom. The first one is related to the modification of
electron dynamics due to existence of the minimum length.
Corresponding correction can be estimated qualitatively with the
use of modified uncertainty relations, (Section II). The second
type of QG correction results from the smearing of Coulomb
potential at the electron radius because the minimum length
effectively implies a non-zero size for the electron. This
correction also can simply be estimated qualitatively along the
Welton's discussion of Lamb shift, (Section III). Both corrections
are of the same order of magnitude.

The third correction is related to the quantum gravity running/reduction
of space-time dimension. The same minimum length gives a clear
physical understanding of space-time dimension reduction.
Physically the existence of minimum length implies the presence of
uncontrollable fluctuations of the background metric as the point
in space-time can not be determined to a better accuracy than
$\delta x_{min}$. Because of this fluctuations four volume also
undergoes fluctuation that under assumption of the four-dimensionality
of the background space-time immediately indicates
an effective reduction of dimension, (Section V). Presently the
concept of QG running/reduction of space-time dimension is no
longer the subject of intuition, it is well established in two
different approaches to QG \cite{AJLLR}. Nevertheless, the present
physical discussion allows us to get a simple analytic expression
of space-time dimension running \cite{mazia1, mazia2}.
Corresponding QG correction to the hydrogen atom can be estimated
by using $D$-dimensional Schr\"{o}dinger (or Dirac) equation for
the Coulomb potential. This type of correction is pretty much of
the same order as the previous ones. For more definiteness we
notice that it is deceptive to count on the suppression of QG
correction due to logarithmic term that appears in the dimension
reduction approach, $1/\ln(l/\delta x_{min})$, because when the QG
corrections become important, that is, when the ratio $l/\delta
x_{min}$ is not extremely large, this term is within the precision
which we can require for QG corrections no matter what the
particular approach is. Let us notice once more that throughout
this paper we neglected the modification of Coulomb potential
$\sim 1/r$ due to dimension reduction $\sim 1/r^{D-2}$.

After determining that the rate of QG corrections due to dimension
reduction are pretty much of the same order as those coming from the
minimum-length modified quantum mechanics, there remains a subtle question
of their signs.
The considerations based on the minimum-length modified quantum
mechanical approach including the qualitative discussions of
sections II and III exhibit the energy shifts with a positive
sign. While, we see that for hydrogen atom as well as harmonic
oscillator energy spectrum the QG correction due to dimension
reduction works with a negative sign. Therefore the sign of total
QG contribution remains obscure.

In a recent paper \cite{DasVagenas} the QG correction to the Lamb
shift due to modified commutation relations was estimated as
\[\delta E \sim E_R \left( {\delta x_{min} \over a} \right)^2\,
{ E_Re^2 \over m\hbar}~.\] It agrees well with the above-found
correction of Lamb shift due to dimension reduction in QED
radiative corrections. We recall that it is by about six orders of
magnitude smaller than the leading QG correction $\sim E_R(\delta
x_{min}/a)^2$. Let us note that the correction to the hydrogen
energy spectrum due to dimension reduction (taking account of the
replacement $1/r \rightarrow  1/r^{D-2}$ as well) was estimated
long ago in \cite{SM} as $\delta E \sim -\varepsilon E_R/6$, which
perfectly agrees with our estimate.

\begin{acknowledgments}

The work of M.\,M. was supported in part by the \emph{CRDF/GRDF} grant
and the \emph{Georgian President Fellowship for Young Scientists}. The
work of Z.\,S. was supported in part by the grants
Sci.\,School-905.\,2006.\,2 and RFBR 06-02-16192-a.

\end{acknowledgments}

\end{document}